\documentstyle[11pt,aaspp4]{article}
\input epsfig.sty
\begin{document}

\title{\bf Broadband spectral analysis of PKS 0528+134: A Report on Six Years of EGRET Observations}

\author{R. Mukherjee} 
\affil{Dept. of Physics \& Astronomy, Barnard College \& Columbia University, New York, NY 10027}

\author {M. B\"ottcher}
\affil{Space Physics \& Astronomy, Rice University, Houston, TX 77005}

\author{R. C. Hartman, P. Sreekumar\footnote{Universities Space Research Association}, \& D. J. Thompson}
\affil{NASA/GSFC, Code 661, Greenbelt, MD 20771}

\author{W. A. Mahoney}
\affil{JPL, California Institute of Technology, Pasadena, CA 91109}

\author{T. Pursimo, A. Sillanp\"a\"a, \& L. O. Takalo}
\affil{Tuorla Observatory, University of Turku, SF-21500, Piikki\"o, Finland}

\begin{abstract}

The multiwavelength spectra of PKS 0528+134 during six years of observations 
by EGRET have been analyzed using synchrotron self-Compton (SSC) and external 
radiation Compton (ERC) models. We find that a two-component model, in which 
the target photons are produced externally to the gamma-ray 
emitting region, but also including an SSC component, is required to 
suitably reproduce the spectral energy distributions of the source. Our 
analysis indicates that there is a trend in the observed properties of 
PKS 0528+134, as the source goes from a gamma-ray low state to a flaring 
state.   We observe that during the higher gamma-ray states, the bulk Lorentz 
factor of the jet increases and the ERC component dominates the high-energy emission. 
Our model calculations indicate the trend that the 
energies of the electrons giving rise to the synchrotron peak decreases, and 
the power-ratio of the gamma-ray and low energy spectral components increases, 
as the source goes from a low to a high gamma-ray state. 

\end{abstract}


\section{Introduction}

PKS 0528+134, a compact radio-loud quasar, is one of the most luminous 
active galactic nuclei (AGN) detected by the Energetic Gamma Ray Experiment 
Telescope (EGRET). The source has been observed several times by EGRET, 
and has also been observed simultaneously at other wavelengths. 
PKS 0528+134 is one of the few gamma-ray blazars that has high flux from 
radio through infrared, 
and has shown extreme variations in the observed gamma-ray emission. The 
polarization and flux density of PKS 0528+134 has been monitored regularly 
at 4.8 GHz, 8.0 GHz, and 14.5 GHz using the Michigan 26-m telescope (Aller 
et al. 1985; Aller \& Aller 1996) and has been found to vary on timescales 
of years. Superluminal motion in PKS 0528+134 was reported 
previously by Pohl et al. (1995). 

Recent findings from geodetic VLBI measurements indicate that outbursts at 
high frequencies can be linked to structural changes in the radio regime 
for PKS 0528+134 and some other blazars (Britzen et al. 1998; Krichbaum et 
al. 1998). VLBI (22-86 GHz) monitoring observations of PKS 0528+134 by 
Krichbaum et al. (1998) show pronounced correlated flux density variations in 
the radio to gamma-ray bands. From their high angular resolution images 
(0.1-0.2 mas), Krichbaum et al. find increasing evidence for a tight 
correlation between this activity and the  production of new jet components. 
Similarly, high time-resolution observations of PKS 0528+134 at 8 GHz from 
the geodetic IRIS campaigns indicate that superluminally moving jet 
components are ejected preceding the time of an observed gamma-ray flare 
(Britzen et al. 1998). 

PKS 0528+134 is faint in the optical with a mean optical brightness of 
$m_v=19.5$ (Wall \& Peacock 1985). The source is located in the Galactic 
anticenter ($l=191.37^\circ$, $b=-11.01^\circ$) behind the diffuse outer 
edge of the dark cloud B30, in the $\lambda$ Ori ring of clouds (Liszt \& 
Wilson 1993; Hogerheijde, et al. 1995) and is heavily absorbed. Estimates 
of Galactic extinction for this source are very uncertain and range from 
$A_v=2.3$ (Liszt \& Wilson 1993) to $A_v=5$, derived from ROSAT observations 
(Zhang et al. 1994). 

There have been few observations of PKS 0528+134 at X-ray energies during 
the time period of the EGRET observations. Observations with ROSAT were 
carried out in 1991 (Zhang et al. 1994) and 1992 (Mukherjee et al. 1996). 
Observations in the medium-hard X-ray range (0.4 - 10 keV) were carried out 
for the first time with ASCA in 1994 and 1995 (Sambruna et al. 1997). 
RXTE observations of PKS 0528+134 were made during 1996 August and September. 
Recently, Ghisellini et al. (1998) reported on the BeppoSAX observations of 
PKS 0528+134 during 1997 February and March. 

At gamma-ray energies, PKS 0528+134 has been detected by both OSSE and 
COMPTEL, besides EGRET. OSSE observed the source to be variable in the 50-150 
keV band (McNaron-Brown et al. 1995) and COMPTEL observations of the source 
above 10 MeV were reported by Collmar et al. (1997). The broadband energy 
spectrum of PKS 0528+134 is dominated by its gamma-ray emission. Recent 
analysis of the gamma-ray spectrum of PKS 0528+134 during the 1993 
observations indicates the presence of a spectral break between the COMPTEL 
and EGRET energies (Collmar et al. 1997). This can be explained by a 
variation of the Doppler beaming factor in the framework of a relativistic 
jet model for blazars (B\"ottcher \& Collmar 1998). 

PKS 0528+134 is one of the four most variable blazars observed by EGRET 
(Mukherjee et al. 1997a) and is one of the few blazars seen by EGRET that has 
flared regularly. Strong variations in flux were observed in 1991, 1993, 1995, 
and 1996. The highest flux from the source was detected in 1993 March, when 
the gamma-ray flux was comparable to that of the Crab. 

The EGRET results have demonstrated that in order to understand the 
production of gamma rays in blazars, it is very important to get truly 
simultaneous coverage across the entire electromagnetic spectrum, before, 
during, and after a flare in the high-energy gamma-ray emission. The high 
gamma-ray luminosity of the blazar suggests that the emission is likely to be 
beamed and, therefore, Doppler-boosted along the line of sight. The jet models 
explain the radio to UV continuum from blazars as synchrotron radiation from 
high energy electrons in a relativistically outflowing jet which has been 
ejected from an accreting supermassive blackhole. The emission in the MeV-GeV 
range  is believed to be due to the inverse Compton scattering of low-energy 
photons by the same relativistic electrons in the jet. However, the source 
of the soft photons that are inverse Compton scattered remains unresolved. 
The soft photons can originate as synchrotron emission either from within 
the jet (``Synchrotron self-Compton'' (SSC) mechanism, e. g. 
Maraschi et al. 1992; Bloom \& Marscher 1996), or 
from a nearby accretion disk (``External radiation Compton '' (ERC) mechanism 
e. g. Dermer \& Schlickeiser 1994), or they can be disk radiation 
reprocessed in broad line region (BLR) clouds. The broad-emission-line 
region in turn can be illuminated either by the disk (Sikora et al. 1994; 
Blandford \& Levinson 1995), or directly by the jet 
(Ghisellini \& Madau 1996). An alternative model by Mannheim (1993) proposes 
that the synchrotron emission is due to the population of both electrons and 
electron-positron pairs arising from shock-accelerated electrons and protons. 
Recently, a model combining the ERC and SSC scenarios has been used to fit the 
simultaneous COMPTEL and EGRET spectra of PKS 0528+134 (B\"ottcher \& Collmar 
1998) measured in Phases 1-3 of the CGRO observations. B\"ottcher \& Collmar 
(1998) suggest that in the case of PKS 0528+134, the SSC process dominates the 
gamma-ray spectrum in the gamma-ray low state, while in the gamma-ray high 
state an additional ERC component becomes dominant at energies above 10 MeV. 

In this article we summarize all the information we have on the 
multiwavelength spectrum of PKS 0528+134 after six years of observations of 
the source with EGRET, and examine the SSC and ERC models as applied to 
PKS 0528+134.  In \S 4 we present model calculations to fit the spectrum of 
PKS 0528+134 during its highest gamma-ray state (1993 March), as well as 
during its lowest observed gamma-ray state (1996 February - March). We 
analyze the multiwavelength spectrum of PKS 0528+134 during its ``flaring,'' 
``intermediate,'' and ``low'' states and test if one consistent model fits 
the spectra during the different states. We start by describing the 
available multiwavelength data on PKS 0528+134 in \S 2. In all the 
multiwaveband data analyzed here, we have used only simultaneous (or nearly 
so) data. Although in several cases the samples are not complete, they do 
provide sufficient information to study the spectral energy distributions of 
PKS 0528+134 in its various states. The model used for the calculations is 
described in \S 3 and we summarize our results in \S 5.

\section{The Data}

\subsection{EGRET Observations}

PKS 0528+134 is one of the 66 blazars detected by EGRET 
(Hunter et al. 1993; Hartman et al. 1999). 
Located close to the Crab and Geminga pulsars, there 
have been several exposures to this source in the past six years of EGRET 
observations. Table 1 lists the dates of observation and the viewing periods 
(VPs) during which the source was within 30$^\circ$ of the EGRET instrument 
axis. A description of the EGRET instrument, standard data processing 
techniques, and the maximum likelihood analysis method 
used to determine the number of source photons are given elsewhere (see for 
example, Hartman et al. (1999), and the references therein). The integrated 
fluxes above 100 MeV, as well as the significances of detection of PKS 0528+134 
is given in Table 1 for the individual VPs. Some VPs were added together 
in order to improve the significance of 
detection. The integrated fluxes from Table 1 are plotted as a function of the 
observation dates in Figure 1. PKS 0528+134 is one of the most variable 
sources seen by EGRET. The source flared to nearly 3 times its mean flux 
level during the 1993 March observations (VP 213). A $\chi^2$ test of the 
data indicates a probability $< 10^{-15}$ that the flux variations are 
consistent with a constant flux. During the Cycle 6 observations (VP 616.1) 
PKS 0528+134 was in one of its lowest gamma-ray states ever and was 
marginally detected at 2.7~$\sigma$ during the month-long exposure to the 
source. 

In order to determine the background-subtracted gamma-ray spectrum of 
PKS 0528+134, the EGRET energy band of 30 MeV -- 10 GeV was divided into 
10 bins, and the number of source photons in each interval was estimated. 
Details of the spectral analysis technique for EGRET sources may be found in 
Nolan et al. (1993). The data were fit to a single power law of the form 
$F(E) = k(E/E_0)^{-\alpha}$ photon cm$^{-2}$ s$^{-1}$ MeV$^{-1}$, where 
$F(E)$ is the flux, $E$ is the energy, $\alpha$ is the photon spectral index, 
$k$ is the coefficient, and $E_0$ is the energy normalization factor. Table 2 
shows the results of the spectral analysis for PKS 0528+134 in those 
VPs chosen here for broadband model fits. For the model fits we have chosen 
a sample of nine VPs with widely varying flux levels.   The classification 
of the gamma-ray state of PKS 0528+134 as ``low,'' ``intermediate,'' or 
``high'' is based on the flux levels and detection significances of the 
source in the different VPs. Although this scheme of classification is some 
what arbitrary, it does enable us to compare the spectra of the source between 
the different observations. In VPs 39, 337, 
and 616.1, the integrated fluxes above 100 MeV from the source were 
the lowest in the sample of VPs chosen for the multiwavelength analysis. 
We designate these observations as ``low'' states. Similarly, we designate 
the observations made in VPs 0.2-0.5, 213, 420, and 528 as 
flaring or ``high'' states. The other VPs analyzed here (VPs 413 and 502) 
are classified as ``intermediate.'' Although there are other VPs 
when the source was detected at a high significance, no broadband spectral 
analysis was done because of poor simultaneous multiwavelength coverage. As 
described further in \S 3, it is important to have observations in the 
optical/IR bands in order to best constrain model fits. 

\subsection{Radio Observations}

Historically PKS 0528+134 has been monitored extensively at radio wavebands. 
In our analysis we 
have only included data that are simultaneous, or nearly so, with the 
EGRET observations. The measurements at 4.8, 8.0, and 14.5 GHz were made with 
the University of Michigan 26 m single dish telescope. The data are available 
on-line from the UMRAO data base (Aller \& Aller 1996). Data 
at 2.7, 4.8, 10.5 and 32 GHz were obtained from the flux density monitoring 
program of extragalactic radio sources with the Effelsberg 100-m telescope 
(Reich et al. 1998). These observations were made using orthogonal 
cross-scans, and the peak flux densities were derived by fitting Gaussian 
functions to the observed data. The Effelsberg 100-m telescope was also used 
to observe the source at 4.85, 10.5 and 23.1 GHz during the EGRET observations 
in VP 616.1. Data at 22 and 37 
GHz were obtained with the 13.7 m Metsahovi Radio Telescope in Finland 
(Ter\"asranta et al. 1992). These observations were made using a dual-beam 
method and were calibrated against DR21 according to the flux density scale 
in Baars et al. (1977). The 90 and 230 GHz 
observations were made with the 15 m Swedish-ESO Submillimeter Telescope 
(SEST) on the European Southern Observatory site of La Sille, Chile. These 
observations are further described in Tornikoski et al. (1996). Observations 
at 150, 273, and 375 GHz were carried out with the 15-m James Clerk Maxwell 
Telescope (JCMT) in Hawaii. Radio 
observations carried out simultaneously with the EGRET measurements in 
VP 616.1 were reported earlier by Mukherjee et al. (1997b). 

\subsection{Optical and Infra-red Observations}

Simultaneous infra-red observations at $J$ (1.25 $\mu$m), $H$ (1.65 $\mu$m), 
and $Ks$ (2.15 $\mu$m) bands were only available 
for the EGRET observations in VP 616.1. These observations
were carried out during the nights of 1997 February 25 and 26 with the
Cassegrain Infrared Camera on the 5-meter telescope at the Palomar Observatory. 
No night-to-night intensity variations were seen to a limit of a few percent. 
Optical observations 
simultaneous with the EGRET observations in VP 213 were made with the 3.5 m 
telescope on Calar Alto, Spain (Wagner et al. 1993; Mukherjee 
et al. 1996). The optical observations of the flare during 1993 March 
were made two days prior to the EGRET observations. 
Three one-minute exposures were taken with an $R$-band filter and a 
Tektronix CCD, and the data were calibrated using a calibration sequence 
of NGC 2419. Observations at the $B$, $V$, and $R$ bands in 1995 October and 
1996 September were made with the 2.5 m Nordic Optical Telescope (NOT), La Palma, 
Canary Islands. A BroCamI CCD and TEK $1024\times1024$ chip with a pixel size 
of 0".176 was used for the measurements. Optical data at $R$ and $I$ bands, during the 
EGRET observations in VP 616.1 were obtained from the figures published in 
Ghisellini et al. (1998). These observations were made at the Torino 
Astronomical Observatory (TAO), with the 1.05 m REOSC astrometric 
telescope equipped with a $1242\times 1152$ pixel CCD camera. TAO observations 
at the $R$ band was also made simultaneous with the EGRET observations in VP 502 
(Villata et al. 1998). The optical monitoring program of the TAO is further 
described by Raiteri et al. (1997). 

The determination of optical extinction for PKS 0528+134 has always been 
difficult. Besides the absorption by the column of neutral Galactic 
hydrogen ($N_H= (2.6\pm0.1)\times 10^{21}$ cm$^{-2}$), absorption due to the 
molecular cloud in the foreground also needs to be taken into account. Here 
we use a value of $N_H= (5.3\pm0.1)\times 10^{21}$ cm$^{-2}$ for the 
total column density, estimated by Ghisellini et al. (1998) using 
archival ASCA observations of the source. This is in good agreement with 
an earlier estimate of $N_H\sim 5\times 10^{21}$ cm$^{-2}$ by Sambruna et al. 
(1997) using 1994 August ASCA data. Much higher extinctions are suggested 
by X-ray observations with ROSAT (Zhang et al. 1994). In their earlier 
analysis of 1993 March data (VP 213), Mukherjee et al. (1996) use a value of 
$N_H \sim 8\times 10^{21}$ cm$^{-2}$, estimated from the ROSAT data. 
However, this value has large uncertainties, and we have re-analyzed the 
broadband spectrum of PKS 0528+134 here using the estimate of 
$N_H$ as determined from the ASCA measurements. The extinction was calculated 
using $A_\lambda/E(B-V)=X$ where $X$ is equal to 
4.1, 3.1, 2.32, 1.50, 0.87, 0.54 and 0.35 for $B$, $V$, $R$, $I$, $J$, 
$H$ and $K$ bands, respectively,   
and $N_H/E(B-V)= 5\times 10^{21}$ cm$^{-2}$ mag$^{-1}$ (Savage \& Mathis 
1979; Rieke \& Labofsky 1985). 
The de-reddened fluxes in Jy were estimated using the relation 
Flux(Jy) $= A \times 10^{-(0.4\times mag)}$, where $A$ is the absolute flux 
density at 0 magnitude, and $mag$ is the corrected magnitude. The values of 
$A$ were obtained from the absolute calibration photometry tables in 
Bessell (1979) (optical) and from Cohen et al. (1992). 

\subsection{X-ray Observations}

X-ray observations of PKS 0528+134, and the corresponding EGRET observations 
(in parentheses) are as follows: ROSAT PSPC observation in 1991 March 
(VPs 0.2-0.5), and 
1992 September (VP 39); ASCA observations in 1994 August (VP 337), and 
1995 February and March (VP 413); RXTE observations in 1996 August and 
September (VP 527 and VP 528); and the BeppoSAX observations in 
1997 February and March (VP 616.1). PKS 0528+134 fluxes at 1 keV for the ROSAT 
observations were obtained from Zhang et al. (1994) and Mukherjee et al. 
(1996). X-ray fluxes for the ASCA observations were obtained from 
the published values in Sambruna et al. (1997). 
We obtained the fluxes for the BeppoSAX 
observations at the 1 keV band from the historical light curves reported in 
Ghisellini et al. (1998). 

 Table 3 shows a list of the observatories contributing to the broad band 
spectral analysis of PKS 0528+134 presented here. 

\section{The Model}

The model used to fit the spectra of PKS 0528+134 is described in detail 
by B\"ottcher, Mause \& Schlickeiser (1997). Here we only summarize the 
salient features and outline the assumptions made. 

The model for 
the geometry of the relativistic AGN jet is as follows. The mass of the 
central black hole is assumed to be $M_{BH}=5\times 10^{10}$ $M_\odot$ and the 
luminosity of the accretion disk is assumed to be $L=5\times 10^{46}$ 
ergs s$^{-1}$. In all our calculations we use a Hubble constant 
$H_0=65$ km s$^{-1}$ Mpc$^{-1}$, $\Lambda=0$, and a $q_0=0.5$ cosmology. 
The model assumes a spherical blob filled with 
ultrarelativistic pair plasma which is moving with a bulk Lorentz factor 
$\Gamma$ along an existing jet structure perpendicular to the accretion disk. 
The angle between the jet axis and the line of sight is $\theta_{obs}$. 
The emerging blob starts out at a height of $z_i$ (injection height) from the 
accretion disk. The particles inside the blob at the injection height 
are distributed isotropically according to 
the power-law $n(\gamma) \propto \gamma^{-s}$, where $\gamma$ is the Lorentz 
factor in the rest frame of the blob, ans $s$ is the spectral index. 
The low and high energy cutoffs in 
the Lorentz factors of the electrons are given by $\gamma_1$ and $\gamma_2$. 
The radio through optical/UV continuum of blazars is explained by 
synchrotron emission from the plasma in relativistic motion.  
The electron-positron pair plasma in the blob cools by synchrotron emission 
and inverse Compton scattering. The model follows the evolution of the 
pair distribution and the 
photon spectra as the blob moves out along the jet. The model takes into 
account synchrotron radiation, SSC and ERC scattering as the various radiative 
energy loss mechanisms in the AGN jet, and photon-photon absorption via 
pair production. In the SSC mechanism, the relativistic electrons 
cool by losing their energy to seed photons originating from the synchrotron 
emission. In the ERC mechanism, the seed photons are external to the jet, 
either directly from the accretion disc, or after being re-scattered by the 
surrounding broad-line region clouds, or both. 
When making comparisons with the 
multiwavelength data, the flux of the blob is time-averaged, since the 
inferred cooling timescale of the plasma in the blob is shorter than the time 
resolution of the $\gamma$-ray observations. 

The model uses a combination 
of SSC and ERC mechanisms to explain the spectral states of PKS 0528+134 
during its various high and low states. A ``two-phase'' model is supported 
by the observations that most gamma-ray blazars go through a ``quiescent phase'' of low 
flux level and a ``flaring phase'' when they have a high gamma-ray flux level 
(Schlickeiser \& Achatz 1992). This has been seen in several cases, e.g. 
PKS 1622-297 (Mattox et al. 1997), 3C 279 (Wehrle et al. 1998), BL Lac 
(Bloom et al. 1997). B\"ottcher \& Collmar (1998) note that a 
homogeneous SSC model (e.g. Bloom \& Marscher 1996) can explain a smooth 
transition of the spectrum from hard X-ray and soft gamma-ray energies to 
a comparatively softer EGRET spectrum. In contrast, in a flaring state, 
the spectral breaks that are observed between the MeV energies and the 
EGRET range can be explained by invoking the ERC mechanism, as such spectral 
breaks cannot be produced by a pure SSC model. A pure version of either model 
is therefore probably an unrealistic simplification as it is quite likely 
that both the SSC and ERC mechanisms are at play during the various states 
of the AGN. 

The Comptel-EGRET $\gamma$-ray spectra of PKS 0528+134 measured in VPs 1, 
213, and 321/322 have been modeled earlier by B\"ottcher \& Collmar (1998). 
The spectral variability of the source at $\gamma$-ray energies was 
explained by these authors by invoking a change in the bulk Lorentz 
factor $\Gamma$ of the blob. It has been shown by Dermer (1995) and 
Dermer, Sturner, \& Schlickeiser (1997) 
that the ERC radiation is related to the Doppler 
factor as $F_{ERC}(\epsilon)\propto D^{3+s}$, where 
$D=(\Gamma[1-\beta_\Gamma\cos\theta_{obs}])^{-1}$, while the SSC radiation has 
a dependence $F_{SSC}(\epsilon)\propto D^{(5+s)/2}.$ Here the bulk Lorentz 
factor $\Gamma$ is given by $\Gamma=(1-\beta_\Gamma)^{-1/2}$. 
Since the ERC mechanism has a stronger 
dependence on the Doppler factor than the SSC component, the difference in 
the spectra between the high and low states was explained by a change in 
the Doppler factor which translates to a change in $\Gamma$. 
In the fits presented here, the model has been modified slightly from the 
version used earlier by B\"ottcher \& Collmar (1998) so that in 
addition to $\Gamma$ the particle distribution of the plasma in the blob is 
also allowed to change between the low and high states.   The model 
parameters that vary are $\Gamma$, the low and high 
electron energy cutoffs, $\gamma_1$ and $\gamma_2$, $n_e$, and $s$. The value of 
the magnetic field $B$ is chosen to be close to the equipartition value. The particles 
are generally less energetic in the high states, when external inverse Compton
scattering becomes more important. This seems plausible since in this case 
Compton losses are much stronger, impeding the acceleration of electrons to 
ultrahigh energies. The additional multiwavelength data at optical, infrared, 
and X-ray energies presented here helps to constrain the model better than 
the fits to the gamma-ray energies alone, as presented by B\"ottcher 
\& Collmar (1998). 

Sambruna et al. (1996) have earlier compared the spectral energy distribution (SED) 
of three complete blazar samples observed by the ROSAT PSPC, namely, RBLs 
(radio-selected BL Lacs), XBLs (X-ray-selected BL Lacs), and FSRQs (flat-spectrum 
radio quasars). A clear separation in the radio-through-X-ray SEDs of RBLs and XBLs 
can be made by defining the objects as LBLs (low-frequency-peaked BL Lacs) or 
HBLs (high-frequency-peaked BL Lacs) on the basis of their X-ray to radio flux 
density. Sambruna et al. (1996) have found that the observed differences in the 
spectral shapes of HBLs, LBLs, and FSRQs correspond to decreasing 
magnetic field densities, 
or decreasing electron densities, or both, and cannot be accounted for by a change 
in the viewing angle solely. In the past we have seen that HBLs can usually be 
fitted with a pure SSC model with very high electron Lorentz factors (e. g. 
Mastichiadis \& Kirk (1997) for Mrk 421; Pian et al. (1998) for Mrk 501), 
while for FSRQs the ERC model generally explains the spectra better 
(e. g. B\"ottcher et al. (1997) for 3C 279). 
Recently B\"ottcher \& Bloom (1998) have shown that BL Lac, an 
example of the LBL class, is indeed intermediate between the HBL and FSRQ classes, 
requiring a relatively weak ERC component. This seems to suggest that generally a 
higher luminosity is related to a stronger external radiation field, implying stronger 
inverse Compton losses. It is possible that this trend is also evident in the different 
states of the same source, as we see in the case of PKS 0528+134 here. 

\section{Multiwavelength Spectral Fits}

In this section we compare the model calculations to the broadband spectrum 
of PKS 0528+134. Several observational results help to constrain the choice 
of parameters in the model fits. VLBI Observations by Pohl et al. (1996) show 
apparent superluminal motion of at least one component of 
$\beta_{\rm app}= 4.4\pm 1.7$ (Pohl et al. 1996),  
implying that we might reasonably expect $\Gamma > 4.4$. 

The accretion disk luminosity $L$ is related to the 
isotropic emission line luminosity, which should be some 
fraction of $L$. 
Sambruna et al. (1997) have found 
$L_{\rm line} =(0.5 - 20)\times10^{46}$ erg s$^{-1}$, from which we 
may conclude that
$L$ should be greater than a few times $10^{46}$ erg s$^{-1}$. The
fact that no big blue bump has been observed in PKS 0528+134
places an upper limit on $L$, which is $\sim 10^{47}$ erg s$^{-1}$. 
An argument for our choice of $M_{BH}$ ($> 10^{10}M_\odot$) here 
is that the disk 
luminosity is probably only a rather small fraction ($< 10$ \%) of the 
Eddington luminosity, because a significant fraction of the accretion power 
goes into the jets. 
This implies that $M_{BH} \gg 10^9 M_{\odot}.$ Although 
the disk luminosity and the black hole mass are not extremely well 
constrained by the modeling procedure, they cannot be too different from the 
values adopted here. 
We estimate that the $L$ and $M_{BH}$ are uncertain by no more than a 
factor of 2.

The size of the emission region is constrained by the observed variability 
of the gamma-ray flux of PKS 0528+134. Short time scale gamma-ray flux 
variations are detectable by EGRET only in the brightest flaring objects, 
due to limited photon statistics. In the case of PKS 0528+134, Mukherjee 
et al. (1996) report a flux variation by a factor of 2 over a time period 
$\tau$ of $\sim 2$ days during the 1993 March flare in the source.  Doubling 
times of $\leq 1$ day have been seen in blazars such as 3C 279 (Wehrle et al. 1998) 
and PKS 1622-297 (Mattox et al. 1997). An upper 
limit for the extent of the emitting region 
is given by $R_B = c\tau D/(1+z)$ where 
$D\equiv \Gamma^{-1}(1-\beta_\Gamma\cos\theta)^{-1}$ is the Doppler 
factor. This gives us an upper limit of $R_B \sim 2\times 10^{15}D$ cm 
for PKS 0528+134. However, shorter variability timescales with 
doubling times of $\leq 0.5$ days would imply a higher $\Gamma$ in all states, and 
the model would have trouble reproducing the observed total luminosity. 

Figures 2 through 10 show the multiwaveband spectra of PKS 0528+134 during 
nine separate EGRET observations of the source. The superimposed solid and 
broken lines are model calculations, appropriate to reproduce the observed 
spectra. In each figure 
the short-dashed line represents the ERC component, the dot-dashed line represents 
the SSC component, and the solid line is the total flux. The long-dashed line 
is the synchrotron spectrum and the dotted line represents the accretion disk 
spectrum. 
In comparing the model parameters between the nine cases studied here, 
we note that the quantities describing the physical parameters of the emitting 
region and the jet are kept the same. The value of $\gamma_2$ is nearly 
the same ($\sim 10^5$) for all the nine spectra studied here. We discuss 
the results of the fits to the individual data sets below. 

The fits to the spectra in the high or flaring gamma-ray states 
(VPs 0.2-0.5, 213, 420, and 528) are shown in Figs. 2, 4, 7, and 9, 
respectively. The fits show that a 
pure SSC model under-predicts the high energy gamma-ray 
spectrum. In the case of VP 213, the spectral break 
found by Collmar et al. (1997) cannot be reproduced by the SSC model, and an 
SSC component with such a high gamma-ray flux would be inconsistent with the 
X-ray spectrum. A pure ERC model, however, reproduces the high energy 
gamma-ray spectrum much better than the SSC model, but underpredicts the 
observed X-ray flux. All the four fits 
require higher bulk Lorentz factors, which are at least 50\% or more 
larger than those for the other cases studied here. The values of $\gamma_1$ 
are lower (the stronger cooling implies that the particles are less energetic) 
than those for the fits to the VPs when the source was not flaring. 
Our choice of $\gamma_1$ at relativistic energies is the reason for the 
X-ray paucity and for the strong low-energy break of the ERC spectrum. A 
similar mechanism could also be achieved by any kind of reacceleration 
mechanism (e.g. Fatuzzo \& Melia 1998). 

The fits to the spectra of PKS 0528+134 in the lowest gamma-ray states 
(VPs 39, 337, and 616.1) are shown in Figs. 3, 5, and 10, 
respectively. In two of the three 
cases the bulk Lorentz factors are the lowest in 
comparison to the other cases. The fits also require a lower particle density 
and higher $\gamma_1$ compared to the other cases.   In the case of VP 337, 
the Comptel data shown in Fig. 5 were not included in the fit. We were unable 
to fit the Comptel data with a model that yielded physically meaningful 
parameters. 

The fits to the spectra measured in VPs 413 and 502, when the gamma-ray fluxes and 
significances of detection are in between the flaring and 
low states, are shown in Figs. 6 and 8, 
respectively. In both the VPs the spectral fits require an intermediate value 
of $\Gamma$, ranging between 7 and 10. 

In all the figures the ERC component appears to exceed the total spectrum 
at high energies. This is because the ERC and SSC components are plotted without taking
into account gamma-gamma absorption. The total spectrum is corrected 
for absorption. This, in fact, shows that gamma-gamma absorption becomes 
important at high photon energies. The double bump in the SSC component 
that appears as a result of the model calculations in the figures 
is primarily due to second-order 
SSC scattering. The two-bump structure in the ERC component might be 
due to the effect of a concave curvature of the evolving electron 
spectrum, which, in turn, is a consequence of Klein-Nishina effects,
leading to less efficient Compton cooling of the highest-energy 
electrons. This is sometimes also reflected in the curved synchrotron 
spectrum ( e. g., in Figs. 3, 5, and 6). 

The fit results are summarized in Table 4. 
Due to the very time-consuming nature of our 
simulations, a detailed parameter study is clearly beyond the scope
of the present paper. However, in order to assess
the sensitivity of our model calculations to variations 
of the parameter values, we plot in Fig. 13 
a set of simulations for each of which a single parameter was changed 
compared to our model
spectrum appropriate for VP 337, as shown in
Fig. 5. All model spectra have been re-normalized
to the measured optical flux. The figure demonstrates
that our model calculations are particularly sensitive 
to changes of the Bulk Lorentz factor $\Gamma$,
the particle spectral index $s$, the low-energy cutoff
$\gamma_1$ of the particle spectrum, and the magnetic
field. In some cases some of the parameters like the pair 
density or $\gamma_2$ are not well constrained due to incomplete frequency 
coverage of the observations. The fit results indicate that a gamma-ray high 
state is primarily related to an increase of the bulk Lorentz factor $\Gamma$. 
Hence the 
ERC component is more dominant during a high gamma-ray state, whereas the SSC 
component provides the bulk of the high-energy emission 
during a low gamma-ray state. Also, the 
particles are generally found to be less energetic in the high states, with 
lower values of $\gamma_1$, which might be related to stronger 
energy losses by external Compton scattering in this state. 

We find that a pure SSC model does not reproduce the spectra well. 
Strong analytical constraints on a pure SSC model fitting the
broadband spectrum of PKS~0528+134 can be deduced from the 
constraints that the intrinsic pair production opacity 
$\tau_{\gamma\gamma} < 1$, and from the total luminosities
in the synchrotron and SSC component, assuming that SSC is
the only relevant radiation process at $\gamma$-ray energies. 
The synchrotron and SSC peak frequencies are related by

$$ \epsilon_{SSC} \approx <\gamma^2> \epsilon_{sy}, \eqno(1) $$
where $\epsilon = h\nu / (m_e c^2)$ is the dimensionless
photon energy, and the implied isotropic luminosities
in both components are

$$ L_{sy} \approx 4 \pi \, R_B^2 \, c \, {u_B}' <\gamma^2> 
\, \tau_T \> D^4, \eqno(2) $$
$$ L_{SSC} \approx <\gamma^2> \, \tau_T \, L_{sy}, \eqno(3) $$
where ${u_B}'$ is the magnetic field energy density in the
blob's comoving frame and $\tau_T = n_e \, R_b \, \sigma_T \ll 1$ 
is the Thomson depth of relativistic electrons in the blob.
Furthermore, we know that

$$ \epsilon_{sy} \approx {B \over B_{cr}} \, {D \over 1 + z} \,
<\gamma^2>, \eqno(4) $$
where $B_{cr} = 4.4 \cdot 10^{13}$~G, and we have the restriction

$$ R_B \le c \, \Delta t_{var} \, {D \over 1 + z}. \eqno(5) $$
from the variability timescale. From Eqs. (1) and (4) we find

$$ {u_B}' \approx {B_{cr}^2 \over 8 \pi} \, {\epsilon_{sy}^4
\over \epsilon_{SSC}^2} \, { (1 + z)^2 \over D^2}. \eqno(6) $$
Thus, Eqs. (2) and (3) yield

$$ D^4 \ge 2 \, {L_{sy}^2 \over L_{SSC}} \, {\epsilon_{SSC}^2 \over 
\epsilon_{sy}^4} \, {1 \over B_{cr}^2 \, c^3 \, (\Delta t_{var})^2}.
\eqno(7) $$
With peak frequencies $\nu_{sy} \approx 3 \cdot 10^{13}$~Hz and
$\nu_{SSC} \approx 10^{20}$~Hz, implied luminosities $L_{sy}
\approx 1.3 \cdot 10^{47}$~erg~s$^{-1}$ and $L_{SSC} \approx
6.3 \cdot 10^{48}$~erg~s$^{-1}$, and a typical variability time
scale of $\sim 2$~days, Eq. (7) gives $D \ge 93$, which appears
to be unreasonably high. Apart from a very high Lorentz factor
$\Gamma \geq 50$, it would require extremely close alignment of
the jet with respect to the line of sight: $\theta_{obs} \le 0.6^o$. 
It would also require the magnetic field to be very weak, 
$B < 0.13$~G (Eq. [4]), implying a very strong deviation from 
equipartition:

$$ {u_B \over u_e} \le {B_{cr}^2 \over 8 \pi \, m_e c^2} {\epsilon_{sy}^{7/2}
\over \epsilon_{SSC}^{3/2}} \, {L_{sy} \over L_{SSC}} \, \sigma_T \, c \,
\Delta t_{var} \approx 0.03. \eqno(8) $$
The estimate (7) is much more severe than the constraint resulting 
from the estimate of the $\gamma\gamma$ opacity, which also applies
to the external inverse-Compton mechanism:

$$ 1 < \tau_{\gamma\gamma} \approx {L_{\gamma} \, \sigma_T \, (1 + z)
\over 4 \pi \, m_e c^4 \, \Delta t_{var} \, D^5}, \eqno(9) $$
which yields $D > 6$ for the values of the observables as quoted
above.

A fit of a pure SSC model to VP 0.2-0.5 (Fig. 11),
using parameters as deduced in the above analytical
estimates shows that (in addition to
the rather implausibly high $\Gamma$ and small
angle $\theta_{obs}$) this model has problems
with the low X-ray flux, and the IR flux 
seems too high. However, the gamma-ray 
spectrum is quite well reproduced by
this calculation. 
%The bar in the optical indicates the typical range of optical fluxes
%for the other high states. 
The model parameters used for this calculation are $\gamma_1 = 2\times10^3$, 
$\gamma_2 = 5\times10^4$, $s = 2.5$, $n_e = 200$ cm$^{-3}$, $B = 0.13$ G, 
$\Gamma = 95$, and $\theta_{obs} = 0.6^\circ$. 
An injection height $z_i=10$ pc $\gg R_B$ was needed in order to suppress 
the ERC component, and is not a very realistic number either. 
Figure 12 shows a pure SSC model calculation to the data in VP 616.1. In this 
case, however, the fit is not much worse than the SSC/ERC model calculation 
shown in Fig. 10. The parameters obtained are $\gamma_1 = 3\times10^3$, 
$\gamma_2 = 2\times10^4$, $s = 2.8$, $n_e = 100$ cm$^{-3}$, $B=3.3$ G, 
$\Gamma = 10$, and $z_i = 1$ pc. 

Sambruna et al. (1997) observe that the relative contributions of the 
SSC and ERC cooling mechanisms depends on the gamma-ray to optical flux ratio 
for a source. They have analyzed the low state data 
of PKS 0528+134 during the 1994 August observations (VP 337), and note that 
even in the low state the gamma-ray flux exceeds the optical flux of the 
source, indicating that the ERC scattering may be the primary cooling 
mechanism.  
Our findings are similar to the results of Sambruna et al. (1997). In all the 
spectra that we have analyzed which have optical observations, the gamma-ray 
flux exceeds the optical flux from the source. This is true even for VP 616.1, 
when the source was in its historically lowest gamma-ray state. Hence it is 
not surprising that there is a significant contribution from the ERC 
scattering to the model fit of the VP 616.1 spectrum. Because of its lower 
gamma-ray to optical flux ratio than for the other VPs studied, the SSC 
mechanism also plays a major role in the cooling of the relativistic 
electrons during VP 616.1. In comparison, in 
VP 213, when the gamma-ray to optical flux ratio was the highest ever, a pure 
SSC model yields a very poor fit to the data, and the contribution of the 
ERC mechanism is significantly larger. 

Recently, Ghisellini et al. (1998) have studied the spectral energy 
distributions of 51 EGRET-detected blazars, and have applied the SSC and ERC 
models to the spectra of these sources. In their analysis the ERC model 
includes contributions from the SSC component, and can therefore be compared 
with our analysis presented here. Although their broadband spectra did not 
include simultaneous measurements at all frequencies, they were able to 
determine trends and correlations among the physical quantities obtained 
from their model calculations. Ghisellini et al. (1998) find evidence for a 
well-defined sequence in the properties of different blazar classes, namely, 
HBL, LBL, 
HPQ (high-polarization quasars), and LPQ (low-polarization quasars). They 
find that the observed spectral properties of these source classes can be 
explained by an increasing contribution of an external radiation field as 
we look at the sequence HBL $\to$ LBL $\to$ HPQ $\to$ LPQ. The trends 
in the observed properties are a decrease in frequencies of the synchrotron 
and inverse Compton peaks, and an increase in the power-ratio of the high and 
low energy spectral components. 
In our analysis here, we see a similar trend in the physical properties of 
the different spectral states of PKS 0528+134. As we study the sequence of 
flaring, intermediate, and low states of the source, we see that the model 
accounts for the flaring state better with a lower $\gamma_1$ (decrease in the 
synchrotron peak) and a higher ratio of synchrotron to gamma-ray luminosity. 
The bulk Lorentz factor for the flaring states is found to be higher than that 
in the non-flaring states, in agreement with the trend found by 
Ghisellini et al. (1998).   The Lorentz factor is an important discriminant 
in the comparisons between the different spectral states, because it 
determines the location of the synchrotron and inverse Compton peaks. 
Although the synchrotron peak for PKS 0528+134 is not well-constrained by 
the data presented here, our conclusions about $\gamma_1$ result primarily 
from the X-ray flux (which is 
SSC emission from low-energy electrons with energies 
around $\gamma_1$) and from the spectral break at MeV energies. 

The model used here explains the broadband spectrum of 
PKS 0528+134 from optical to high energy gamma rays reasonably well, but 
underpredicts the radio flux. This is because the low energy cutoff in the 
electron energy produces a sharp cutoff in the synchrotron spectrum 
below $\sim 10^{13}$ Hz where 
the blob becomes optically thick to synchrotron-self 
absorption. The radio flux only varies with very small amplitude. Thus it is 
most probably produced in the outer jet regions which are less 
influenced by single blob injection events. The outer jet regions, where 
the kinetic power of the jet is dissipated and the expansion 
of the jet becomes relevant, are not included in our simulations.

The B\"ottcher \& Collmar (1998) model generally has problems with the 
hard optical spectra which resemble more the instantaneous rather than the 
time-averaged synchrotron spectra of the evolving electron population. 
However, recently Chiaberge \& Ghisellini (1998) have noted that even 
if the electron cooling time scale is much shorter than the light travel time 
through the blob, no strong variability results on a time scale shorter than 
the light travel time. Thus it is likely that we are seeing the time-averaged 
spectrum even in the optical. Alternatively, it is possible that 
reacceleration at the low-energy end of the electron spectrum during the early 
stages of the blob evolution would lead to an optical spectrum resembling the 
instantaneous spectrum of the source. 

\section{Summary \& Conclusions}

We present model calculations of the multiwavelength spectrum of 
PKS 0528+134 during its various gamma-ray states. 
The broad-band spectrum of PKS 0528+134 is dominated by the emission in 
gamma rays. This is true even when the source was in its lowest ever gamma-ray 
state. 

We find that the spectral energy distribution of PKS 0528+134 can be modeled 
as follows: the radio to UV emission can be explained as synchrotron 
emission from relativistic electrons in a uniform relativistically moving 
plasma. The high energy 
emission is due to the inverse Compton scattering of seed photons off the 
relativistic electrons. The electron cooling process is a combination of SSC 
and ERC mechanisms, based on the different dependences of the ERC and the SSC 
components on the Doppler boosting factor. In the gamma-ray high state, due 
to an increase in the bulk Lorentz factor of the jet, the ERC component 
becomes the only significant part of the gamma-ray spectrum. 

The model that yields an acceptable fit to the data assumes the mass of the 
central black hole $M_{BH}=5\times 10^{10}$ $M_\odot$ and the luminosity of 
the accretion disk to be $L=5\times 10^{46}$ ergs s$^{-1}$ for PKS 0528+134. 

It has generally been seen that sources with higher bolometric luminosity, 
dominated more strongly by the gamma-ray luminosity, 
are fitted better with the ERC model rather than a pure SSC model. 
Ghisellini et al. (1998) find that the properties of HBL, LBL, and FSRQs are 
located along a sequence, with the HBL characterized by the lowest intrinsic 
power and weakest external radiation field.   Our model calculations indicate 
that a similar trend exists in the flaring and low spectral states of 
PKS 0528+134. 
It would be interesting to test if this is generally true by analyzing the 
flaring and low-state spectra of other blazars detected by EGRET that 
have exhibited similar extreme variations in their gamma-ray flux (e. g. 
3C 279, PKS 1622-297). 

The spectral variability in PKS 0528+134 appears to arise from the different 
Doppler boosting patterns of the SSC and the ERC radiations. 
The relative contributions of the SSC and ERC cooling mechanisms seem 
to be related to the optical to gamma-ray flux ratio from the source. The 
SSC mechanism plays a larger role if the source is in a low flux state. The 
ERC mechanism is the dominant cooling mechanism when the source is in a 
high gamma-ray state. 

  The main source of uncertainty in the model parameters arises from 
incomplete spectral coverage of PKS 0528+134 during the various observation 
periods. 
More precise (contemporaneous) measurements of the high-energy cutoffs 
of the synchrotron (in the UV) and gamma-ray spectral components are 
necessary to constrain models better, especially parameters like the 
high-energy cutoff of the electron spectrum. Future observations of AGN 
with successor gamma-ray observatories like INTEGRAL (e.g. Lichti et al. 1996) 
and GLAST (e.g. Michelson 1996) should play a key role in resolving the 
physics of these powerful sources. 

\bigskip

R. Mukherjee acknowledges support by NASA Grant NAG5-3696. 
This research has made use of data from the University of Michigan Radio
Astronomy Observatory which is supported by the National Science Foundation 
and by funds from the University of Michigan. M. B\"ottcher acknowledges 
support by NASA Grant NAG5-4055. R. Mukherjee acknowledges many useful 
discussions with J. Halpern. The authors would like to thank the 
anonymous referee for the constructive comments on the paper.

\clearpage

\begin{deluxetable}{llccc}
\tablenum{1}
\tablewidth{40pc}
\tablecaption{EGRET observations of PKS 0528+134 from 1991 to 1997}
\tablehead{
\colhead{Viewing}      & \colhead{Observation}     & \colhead{Flux $\times10^{-7}$ } & \colhead{Significance} & \colhead{Inclination Angle} \\
\colhead{Period}      & \colhead{Dates} & \colhead{ph cm$^{-2}$ s$^{-1}$ } & \colhead {$\sigma$}& \colhead {of Source}}

\startdata
 0.2-0.5        &1991 Apr 22 -- May 07  &$12.9\pm0.9$   &20.3   &$8.^\circ0$    \nl
 1.0            &1991 May 16 -- 30      &$8.5\pm0.8$    &13.5   &$6.^\circ3$    \nl
 2.1            &1991 Jun 08 -- 15      &$3.6\pm1.1$    &3.8    &$5.^\circ1$    \nl
 36.0+36.5      &1992 Aug 11 -- 20      &$< 5.5^a$      & -     &$22.^\circ0$   \nl
 39.0           &1992 Sep 01 -- 17      &$3.2\pm1.4$    &2.6    &$23.^\circ9$   \nl
 213.0          &1993 Mar 23 -- 29      &$30.8\pm3.5$   &13.6   &$9.^\circ1$    \nl
 221.0          &1993 May 13 -- 24      &$2.3\pm1.2$    &2.2    &$6.^\circ4$    \nl 
 310.0          &1993 Dec 01 -- 13      &$\rm <4.0^a$   & -     &$15.^\circ7$   \nl 
 321.1+321.5    &1994 Feb 08 -- 17      &$4.9\pm1.2$    &5.0    &$12.^\circ9$   \nl 
 337.0          &1994 Aug 09 -- 29      &$3.2\pm1.0$    &3.6    &$13.^\circ5$   \nl
 412.0          &1995 Feb 28 -- Mar 07  &$9.1\pm2.1$    &5.8    &$13.^\circ1$   \nl
 413.0          &1995 Mar 07 -- 21      &$9.0\pm1.3$    &9.3    &$7.^\circ7$    \nl
 419.1          &1995 Apr 04 -- 11      &$12.1\pm2.4$   &6.6    &$17.^\circ4$   \nl
 419.5          &1995 May 09 -- 23      &$12.0\pm2.2$   &7.5    &$20.^\circ9$   \nl
 420.0          &1995 May 23 -- Jun 06  &$13.0\pm1.6$   &11.2   &$9.^\circ8$    \nl
 426.0          &1995 Aug 08 -- 22      &$5.5\pm1.7$    &4.0    &$8.^\circ5$    \nl
 502.0          &1995 Oct 17 -- 31      &$5.7\pm0.8$    &8.5    &$0.^\circ9$    \nl
 526.0          &1996 Jul 30 -- Aug 13  &$5.2\pm1.2$    &5.4    &$8.^\circ3$    \nl
 527.0          &1996 Aug 13 -- 20      &$5.1\pm1.6$    &4.3    &$9.^\circ3$    \nl
 528.0          &1996 Aug 20 -- 27      &$17.3\pm2.1$   &11.7   &$12.^\circ1$   \nl
 616.1           &1997 Feb 18 -- Mar 18  &$1.1\pm0.5$    &2.7    &$0.^\circ0$    \nl

\tablenotetext{}{$ \rm a$: $2\sigma$ upper limit}
\enddata
\end{deluxetable}

\begin{deluxetable}{llccc}
\tablenum{2}
\tablewidth{26pc}
\tablecaption{EGRET spectral analysis for PKS 0528+134 in selected observations}
\tablehead{
\colhead{Viewing}      & \colhead{Spectral}     & \colhead{k $\times10^{-9}$ } & \colhead{$E_0$} & \colhead{$\chi^2/n_f$} \\
\colhead{Period}      & \colhead{Index ($\alpha$)} & \colhead{ph cm$^{-2}$ s$^{-1}$ MeV$^{-1}$} & \colhead {MeV}& \colhead { }\\
}
\startdata
 0.2-0.5        &$2.27\pm0.07$          &$3.26\pm0.19$  & 199   & 1.53 \nl
  39.0          &$2.39\pm0.78$          &$1.17\pm0.61$  & 166   & 0.13 \nl
 213.0          &$2.21\pm0.11$          &$8.24\pm0.76$  & 200   & 0.78 \nl
 337.0          &$2.68\pm0.44$          &$2.15\pm0.57$  & 135   & 0.52 \nl
 413.0          &$ 2.21 \pm 0.16$       &$0.92\pm0.15$  & 255   & 1.07 \nl
 420.0          &$ 2.37 \pm 0.13$       &$2.48\pm0.29$  & 206   & 1.19 \nl
 502.0          &$2.32\pm0.16$          &$1.16\pm0.17$  & 235   & 0.55 \nl
 528.0          &$ 2.44 \pm 0.44$       &$2.73\pm0.76$  & 159   & 0.19 \nl
 616.1           &$ 2.51\pm 0.47$        &$0.22\pm0.10$  & 233   & 0.76 \nl
\enddata
\end{deluxetable}

\begin{deluxetable}{llll}
\tablenum{3}
\tablewidth{34pc}
\tablecaption{Contributing observatories and spectral coverage}
\tablehead{
\colhead{Observatory} & \colhead{Instrument/Telescope} & \colhead{Spectral region} & \colhead{band}\\}
\startdata
CGRO        & EGRET  & $\gamma$-ray & 0.03 - 10 GeV          \nl
            & COMPTEL& $\gamma$-ray & 0.75 - 30 MeV          \nl
ROSAT       & PSPC   & X-ray        & 0.5  -  2 keV          \nl
ASCA        &        & X-ray        & 0.5  -  2 keV          \nl
Beppo SAX   &        & X-ray        & 0.5  -  2 keV          \nl
Palomar     & 5 m    & infra red    & $J$ $H$ $K$            \nl
Torino      & 1.05 m & optical      & $R$ $I$                \nl
NOT         & 2.56 m & optical      & $B$ $V$ $R$            \nl
Calar Alto  & 3.5 m  & optical      & $R$                    \nl
%Vainu Bappu & 2.34 m & optical      & $V$                    \nl
Effelsberg  & 100 m  & mm/cm        & 2.69, 4.8, 10.5, 23.1, 32 GHz\nl
SEST        & 15 m   & mm           & 90, 230 GHz            \nl
Mets\"ahovi & 13.7 m & mm/cm        & 22, 36.8 GHz           \nl
JCMT        & 15 m   & mm/sub-mm    & 150, 273, 375 GHz      \nl
UMRAO       & 26 m   & cm           & 4.8, 8.0, 14.5 GHz     \nl
\enddata
\end{deluxetable}

\begin{deluxetable}{llcccccl}
\tablenum{4}
\tablewidth{29pc}
\tablecaption{Interesting model parameters for the different viewing periods}
\tablehead{
\colhead{VP}& \colhead{$\gamma_1$} & \colhead{$\gamma_2$ } & \colhead{$s$} & \colhead{$\Gamma$} & $n_e$ & $B$ & Gamma-ray\\
\colhead{} & \colhead{} & \colhead{} & \colhead{} & \colhead{} & \colhead{cm$^{-3}$} & \colhead{G} & State$^*$\\
}
\startdata
  0.2-0.5  &180   & $10^5$          & 2.5  & 15 & 290 & 2.5 & High\nl 
  39.0     &1000  & $10^5$          & 2.5  &  5 & 150 & 2.5 & Low\nl
 213.0     &120   & $6\times 10^4$  & 2.6  & 20 & 150 & 1.0 & High\nl
 337.0     &900   & $10^5$          & 2.5  & 10 & 150 & 3.0 & Low\nl 
 413.0     &1000  & $        10^5$  & 2.5  & 10 & 180 & 2.0 & Intermediate\nl
 420.0     &500   & $7\times 10^4$  & 2.5  & 20 & 180 & 2.0 & High\nl
 502.0     &1000  & $10^5$          & 2.5  &  7 & 180 & 3.2 & Intermediate\nl
 528.0     &600   & $        10^5$  & 2.2  & 10 & 180 & 2.5 & High\nl
 616.1     &1000  & $        10^5$  & 2.5  &  5 &  50 & 2.5 & Low\nl
\tablenotetext{}{$*$ The high energy gamma-ray state of the source 
(EGRET energy range), based on the criteria described in the text. }
\enddata
\end{deluxetable}

\clearpage
{\bf Figure Captions}
\bigskip

Figure 1: Flux of gamma rays above 100 MeV from PKS 0528+134 over the period 1991 April 
to 1997 March (Phases 1 through Cycle 6). $2\sigma$ upper limits are shown as 
downward arrows. 

\smallskip
Figure 2: Fit to the spectral energy distribution of PKS 0528+134 in 
VP 0.2-0.5. The long-dashed line represents the synchrotron spectrum, the short-dashed 
line is the ERC component, the dot-dashed line is the 
SSC component, and the solid line represents the combined SSC + ERC model. 
The dotted line is the accretion disk spectrum. 
Parameters: $\gamma_1 = 180        $, 
$\gamma_2 =        10^5$, $s = 2.5$, $n_e = 290$ cm$^{-3}$, $B = 2.5 $ G, 
$\Gamma = 15$.

\smallskip
Figure 3: Fit to the spectral energy distribution of PKS 0528+134 in 
VP 39. The long-dashed line represents the synchrotron spectrum, the short-dashed 
line is the ERC component, the dot-dashed line is the 
SSC component, and the solid line represents the combined SSC + ERC model. 
The dotted line is the accretion disk spectrum. 
Parameters: $\gamma_1 =        10^3$, 
$\gamma_2 =        10^5$, $s = 2.5$, $n_e = 150$ cm$^{-3}$, $B = 2.5 $ G, 
$\Gamma =  5$.

\smallskip
Figure 4: Fit to the spectral energy distribution of PKS 0528+134 in 
VP 213. The long-dashed line represents the synchrotron spectrum, the short-dashed 
line is the ERC component, the dot-dashed line is the 
SSC component, and the solid line represents the combined SSC + ERC model. 
The dotted line is the accretion disk spectrum. 
Parameters: $\gamma_1 = 120        $, 
$\gamma_2 = 6\times10^4$, $s = 2.6$, $n_e = 150$ cm$^{-3}$, $B = 1.0 $ G, 
$\Gamma = 20$.

\smallskip
Figure 5: Fit to the spectral energy distribution of PKS 0528+134 in 
VP 337. The long-dashed line represents the synchrotron spectrum, the short-dashed 
line is the ERC component, the dot-dashed line is the 
SSC component, and the solid line represents the combined SSC + ERC model. 
The dotted line is the accretion disk spectrum. 
Parameters: $\gamma_1 = 900$, 
$\gamma_2 =        10^5$, $s = 2.5$, $n_e = 150$ cm$^{-3}$, $B = 3.0 $ G, 
$\Gamma = 10$.

\smallskip
Figure 6: Fit to the spectral energy distribution of PKS 0528+134 in 
VP 413. The long-dashed line represents the synchrotron spectrum, the short-dashed 
line is the ERC component, the dot-dashed line is the 
SSC component, and the solid line represents the combined SSC + ERC model. 
The dotted line is the accretion disk spectrum. 
Parameters: $\gamma_1 = 1000       $, 
$\gamma_2 =        10^5$, $s = 2.5$, $n_e = 180$ cm$^{-3}$, $B = 2.0 $ G, 
$\Gamma = 10$.

\smallskip
Figure 7: Fit to the spectral energy distribution of PKS 0528+134 in 
VP 420. The long-dashed line represents the synchrotron spectrum, the short-dashed 
line is the ERC component, the dot-dashed line is the 
SSC component, and the solid line represents the combined SSC + ERC model. 
The dotted line is the accretion disk spectrum. 
Parameters: $\gamma_1 = 500        $, 
$\gamma_2 = 7\times10^4$, $s = 2.5$, $n_e = 180$ cm$^{-3}$, $B = 2.0 $ G, 
$\Gamma = 20$.

\smallskip
Figure 8: Fit to the spectral energy distribution of PKS 0528+134 in 
VP 502. The long-dashed line represents the synchrotron spectrum, the short-dashed 
line is the ERC component, the dot-dashed line is the 
SSC component, and the solid line represents the combined SSC + ERC model. 
The dotted line is the accretion disk spectrum. 
Parameters: $\gamma_1 = 1000       $, 
$\gamma_2 =        10^5$, $s = 2.5$, $n_e = 180$ cm$^{-3}$, $B = 3.2 $ G, 
$\Gamma =  7$.

\smallskip
Figure 9: Fit to the spectral energy distribution of PKS 0528+134 in 
VP 528. The long-dashed line represents the synchrotron spectrum, the short-dashed 
line is the ERC component, the dot-dashed line is the 
SSC component, and the solid line represents the combined SSC + ERC model. 
The dotted line is the accretion disk spectrum. 
Parameters: $\gamma_1 = 600        $, 
$\gamma_2 =        10^5$, $s = 2.2$, $n_e = 180$ cm$^{-3}$, $B = 2.5 $ G, 
$\Gamma = 10$.

\smallskip
Figure 10: Fit to the spectral energy distribution of PKS 0528+134 in 
VP 616.1. The long-dashed line represents the synchrotron spectrum, the short-dashed 
line is the ERC component, the dot-dashed line is the 
SSC component, and the solid line represents the combined SSC + ERC model. 
The dotted line is the accretion disk spectrum. 
Parameters: $\gamma_1 = 1000       $, 
$\gamma_2 =        10^5$, $s = 2.5$, $n_e =  50$ cm$^{-3}$, $B = 2.5 $ G, 
$\Gamma =  5$.

\smallskip
Figure 11: Attempt to fit to the spectral energy distribution of PKS 0528+134 
in VP 0.2-0.5 with a pure SSC model. The long-dashed line represents the synchrotron spectrum, the short-dashed 
line is the ERC component, the dot-dashed line is the 
SSC component, and the solid line represents the combined SSC + ERC model. 
The dotted line is the accretion disk spectrum. 
Parameters: $\gamma_1 = 2\times10^3$, 
$\gamma_2 = 5\times10^4$, $s = 2.5$, $n_e = 200$ cm$^{-3}$, $B = 0.13$ G, 
$\Gamma = 95$, $\theta_{obs} = 0.6^\circ$, and $z_i=10$ pc.

\smallskip
Figure 12: Attempt to fit to the spectral energy distribution of PKS 0528+134 in 
VP 616.1 with a pure SSC model. The long-dashed line represents the synchrotron spectrum, the short-dashed 
line is the ERC component, the dot-dashed line is the 
SSC component, and the solid line represents the combined SSC + ERC model. 
The dotted line is the accretion disk spectrum. 
Parameters: $\gamma_1 = 3\times10^3$, 
$\gamma_2 = 2\times10^4$, $s = 2.8$, $n_e = 100$ cm$^{-3}$, $B=3.3$ G, 
$\Gamma = 10$, and $z_i = 1$ pc.

\smallskip
Figure 13: Fits to the spectral energy distribution of PKS 0528+134 in 
VP 337, in which a single parameter is changed compared to our 
model spectrum appropriate for VP 337, as shown in Fig. 5. The figure is an 
indicator of the sensitivity of our model calculations to variations of the 
parameter space.

\clearpage

\begin{figure}[t!] % fig 1
\centerline{\epsfig{file=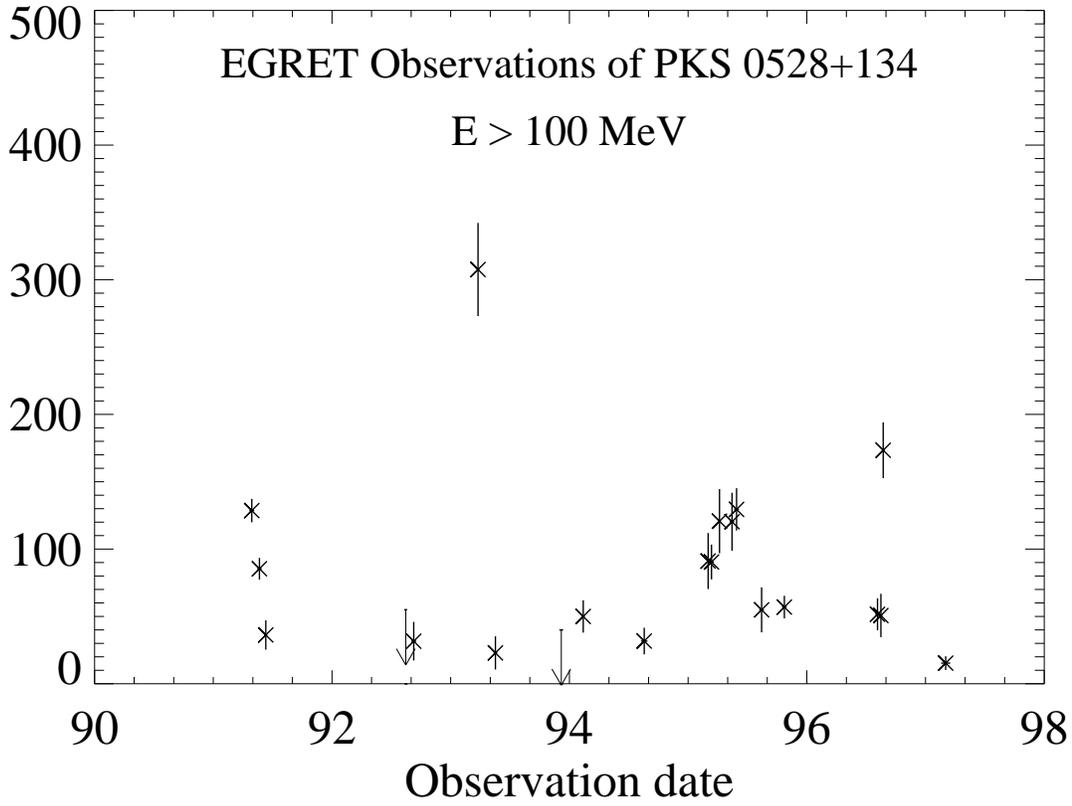,height=5.0in,bbllx=100pt,bblly=300pt,bburx=650pt,bbury=700pt,clip=.}}
\vspace{10pt}
\caption{Flux of gamma rays above 100 MeV from PKS 0528+134 over the period 1991 April 
to 1997 March (Phases 1 through Cycle 6). $2\sigma$ upper limits are shown as 
downward arrows.}
\label{fig1}
\end{figure}
 
\begin{figure}[t!] % fig 2
\centerline{\epsfig{file=mukherjeer_fig2.ps,height=5.0in,bbllx=50pt,bblly=-600pt,bburx=700pt,bbury=0pt,clip=.}}
\vspace{10pt}
\caption{Fit to the spectral energy distribution of PKS 0528+134 in 
VP 0.2-0.5. The long-dashed line represents the synchrotron spectrum, the short-dashed 
line is the ERC component, the dot-dashed line is the 
SSC component, and the solid line represents the combined SSC + ERC model. 
The dotted line is the accretion disk spectrum. 
Parameters: $\gamma_1 = 180        $, 
$\gamma_2 =        10^5$, $s = 2.5$, $n_e = 290$ cm$^{-3}$, $B = 2.5 $ G, 
$\Gamma = 15$.}                                               
\label{fig2}
\end{figure}
 
\begin{figure}[t!] % fig 3
\centerline{\epsfig{file=mukherjeer_fig3.ps,height=5.0in,bbllx=50pt,bblly=-600pt,bburx=700pt,bbury=0pt,clip=.}}
\vspace{10pt}
\caption{Fit to the spectral energy distribution of PKS 0528+134 in 
VP 39. The long-dashed line represents the synchrotron spectrum, the short-dashed 
line is the ERC component, the dot-dashed line is the 
SSC component, and the solid line represents the combined SSC + ERC model. 
The dotted line is the accretion disk spectrum. 
Parameters: $\gamma_1 =        10^3$, 
$\gamma_2 =        10^5$, $s = 2.5$, $n_e = 150$ cm$^{-3}$, $B = 2.5 $ G, 
$\Gamma =  5$.}                                              
\label{fig3}
\end{figure}
 
\begin{figure}[t!] % fig 4
\centerline{\epsfig{file=mukherjeer_fig4.ps,height=5.0in,bbllx=50pt,bblly=-600pt,bburx=700pt,bbury=0pt,clip=.}}
\vspace{10pt}
\caption{Fit to the spectral energy distribution of PKS 0528+134 in 
VP 213. The long-dashed line represents the synchrotron spectrum, the short-dashed 
line is the ERC component, the dot-dashed line is the 
SSC component, and the solid line represents the combined SSC + ERC model. 
The dotted line is the accretion disk spectrum. 
Parameters: $\gamma_1 = 120        $, 
$\gamma_2 = 6\times10^4$, $s = 2.6$, $n_e = 150$ cm$^{-3}$, $B = 1.0 $ G, 
$\Gamma = 20$.}
\label{fig4}
\end{figure}
 
\begin{figure}[t!] % fig 5
\centerline{\epsfig{file=mukherjeer_fig5.ps,height=5.0in,bbllx=50pt,bblly=-600pt,bburx=700pt,bbury=0pt,clip=.}}
\vspace{10pt}
\caption{Fit to the spectral energy distribution of PKS 0528+134 in 
VP 337. The long-dashed line represents the synchrotron spectrum, the short-dashed 
line is the ERC component, the dot-dashed line is the 
SSC component, and the solid line represents the combined SSC + ERC model. 
The dotted line is the accretion disk spectrum. 
Parameters: $\gamma_1 = 900$, 
$\gamma_2 =        10^5$, $s = 2.5$, $n_e = 150$ cm$^{-3}$, $B = 3.0 $ G, 
$\Gamma = 10$.}
\label{fig5}
\end{figure}
 
\bigskip
\begin{figure}[t!] % fig 6
\centerline{\epsfig{file=mukherjeer_fig6.ps,height=5.0in,bbllx=50pt,bblly=-600pt,bburx=700pt,bbury=0pt,clip=.}}
\vspace{10pt}
\caption{Fit to the spectral energy distribution of PKS 0528+134 in 
VP 413. The long-dashed line represents the synchrotron spectrum, the short-dashed 
line is the ERC component, the dot-dashed line is the 
SSC component, and the solid line represents the combined SSC + ERC model. 
The dotted line is the accretion disk spectrum. 
Parameters: $\gamma_1 = 1000       $, 
$\gamma_2 =        10^5$, $s = 2.5$, $n_e = 180$ cm$^{-3}$, $B = 2.0 $ G, 
$\Gamma = 10$.}
\label{fig6}
\end{figure}
 
\bigskip
\begin{figure}[t!] % fig 7
\centerline{\epsfig{file=mukherjeer_fig7.ps,height=5.0in,bbllx=50pt,bblly=-600pt,bburx=700pt,bbury=0pt,clip=.}}
\vspace{10pt}
\caption{Fit to the spectral energy distribution of PKS 0528+134 in 
VP 420. The long-dashed line represents the synchrotron spectrum, the short-dashed 
line is the ERC component, the dot-dashed line is the 
SSC component, and the solid line represents the combined SSC + ERC model. 
The dotted line is the accretion disk spectrum. 
Parameters: $\gamma_1 = 500        $, 
$\gamma_2 = 7\times10^4$, $s = 2.5$, $n_e = 180$ cm$^{-3}$, $B = 2.0 $ G, 
$\Gamma = 20$.}
\label{fig7}
\end{figure}
 
\begin{figure}[t!] % fig 8
\centerline{\epsfig{file=mukherjeer_fig8.ps,height=5.0in,bbllx=50pt,bblly=-600pt,bburx=700pt,bbury=0pt,clip=.}}
\vspace{10pt}
\caption{Fit to the spectral energy distribution of PKS 0528+134 in 
VP 502. The long-dashed line represents the synchrotron spectrum, the short-dashed 
line is the ERC component, the dot-dashed line is the 
SSC component, and the solid line represents the combined SSC + ERC model. 
The dotted line is the accretion disk spectrum. 
Parameters: $\gamma_1 = 1000       $, 
$\gamma_2 =        10^5$, $s = 2.5$, $n_e = 180$ cm$^{-3}$, $B = 3.2 $ G, 
$\Gamma =  7$.}
\label{fig8}
\end{figure}
 
\bigskip
\begin{figure}[t!] % fig 9
\centerline{\epsfig{file=mukherjeer_fig9.ps,height=5.0in,bbllx=50pt,bblly=-600pt,bburx=700pt,bbury=0pt,clip=.}}
\vspace{10pt}
\caption{Fit to the spectral energy distribution of PKS 0528+134 in 
VP 528. The long-dashed line represents the synchrotron spectrum, the short-dashed 
line is the ERC component, the dot-dashed line is the 
SSC component, and the solid line represents the combined SSC + ERC model. 
The dotted line is the accretion disk spectrum. 
Parameters: $\gamma_1 = 600        $, 
$\gamma_2 =        10^5$, $s = 2.2$, $n_e = 180$ cm$^{-3}$, $B = 2.5 $ G, 
$\Gamma = 10$.}
\label{fig9}
\end{figure}
 
\bigskip
\begin{figure}[t!] % fig 10
\centerline{\epsfig{file=mukherjeer_fig10.ps,height=5.0in,bbllx=50pt,bblly=-600pt,bburx=700pt,bbury=0pt,clip=.}}
\vspace{10pt}
\caption{Fit to the spectral energy distribution of PKS 0528+134 in 
VP 616.1. The long-dashed line represents the synchrotron spectrum, the short-dashed 
line is the ERC component, the dot-dashed line is the 
SSC component, and the solid line represents the combined SSC + ERC model. 
The dotted line is the accretion disk spectrum. 
Parameters: $\gamma_1 = 1000       $, 
$\gamma_2 =        10^5$, $s = 2.5$, $n_e =  50$ cm$^{-3}$, $B = 2.5 $ G, 
$\Gamma =  5$.}
\label{fig10}
\end{figure}
 
\bigskip
\begin{figure}[t!] % fig 11
\centerline{\epsfig{file=mukherjeer_fig11.ps,height=5.0in,bbllx=50pt,bblly=-600pt,bburx=700pt,bbury=0pt,clip=.}}
\vspace{10pt}
\caption{Attempt to fit to the spectral energy distribution of PKS 0528+134 
in VP 0.2-0.5 with a pure SSC model. The long-dashed line represents the synchrotron spectrum, the short-dashed 
line is the ERC component, the dot-dashed line is the 
SSC component, and the solid line represents the combined SSC + ERC model. 
The dotted line is the accretion disk spectrum. 
Parameters: $\gamma_1 = 2\times10^3$, 
$\gamma_2 = 5\times10^4$, $s = 2.5$, $n_e = 200$ cm$^{-3}$, $B = 0.13$ G, 
$\Gamma = 95$, $\theta_{obs} = 0.6^\circ$, and $z_i=10$ pc. }
\label{fig11}
\end{figure}

\bigskip
\begin{figure}[t!] % fig 12
\centerline{\epsfig{file=mukherjeer_fig12.ps,height=5.0in,bbllx=50pt,bblly=-600pt,bburx=700pt,bbury=0pt,clip=.}}
\vspace{10pt}
\caption{Attempt to fit to the spectral energy distribution of PKS 0528+134 in 
VP 616.1 with a pure SSC model. The long-dashed line represents the synchrotron spectrum, the short-dashed 
line is the ERC component, the dot-dashed line is the 
SSC component, and the solid line represents the combined SSC + ERC model. 
The dotted line is the accretion disk spectrum. 
Parameters: $\gamma_1 = 3\times10^3$, 
$\gamma_2 = 2\times10^4$, $s = 2.8$, $n_e = 100$ cm$^{-3}$, $B=3.3$ G, 
$\Gamma = 10$, and $z_i = 1$ pc. }
\label{fig12}
\end{figure}
 
\bigskip
\begin{figure}[t!] % fig 13
\centerline{\epsfig{file=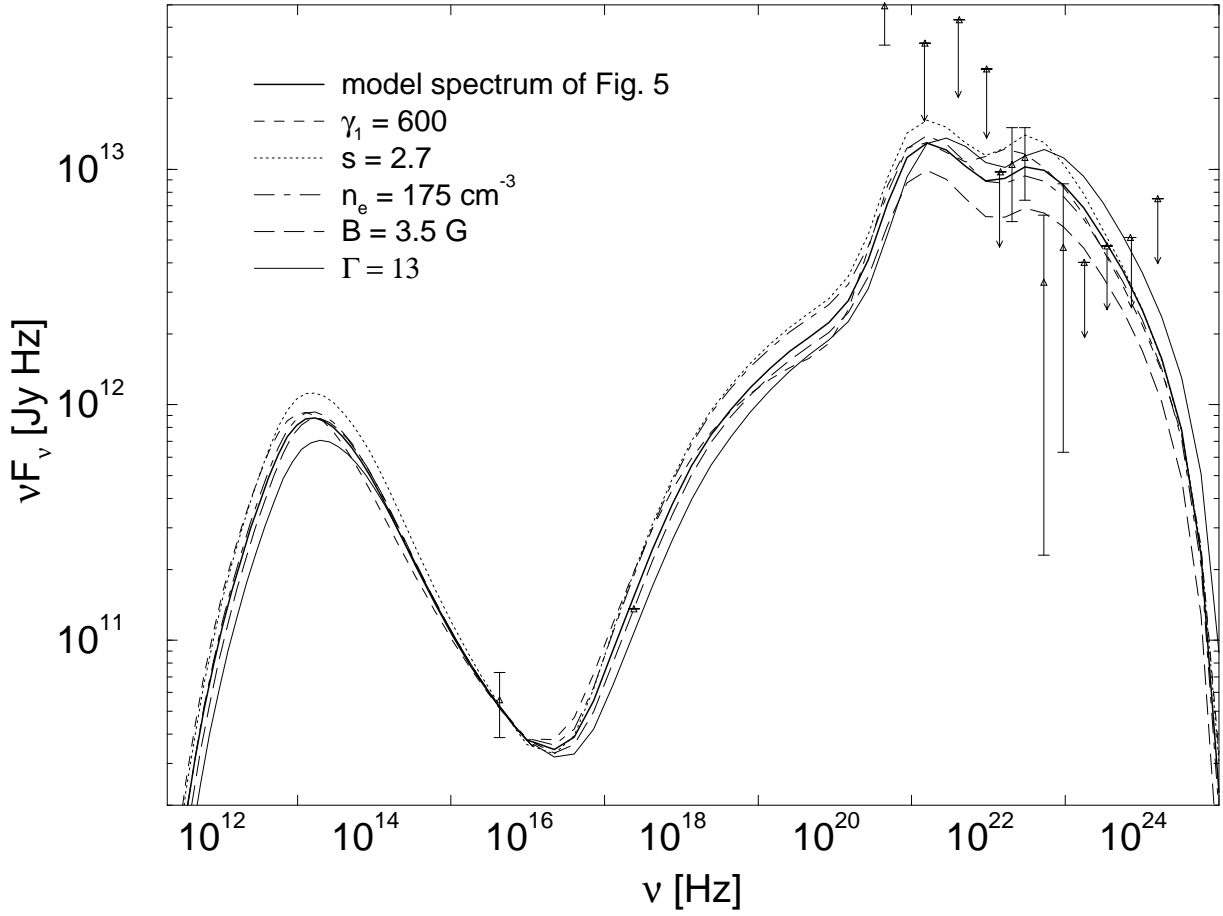,height=5.0in,bbllx=35pt,bblly=85pt,bburx=670pt,bbury=565pt,clip=.}}
\vspace{10pt}
\caption{Fits to the spectral energy distribution of PKS 0528+134 in 
VP 337, in which a single parameter is changed compared to our 
model spectrum appropriate for VP 337, as shown in Fig. 5. The figure is an 
indicator of the sensitivity of our model calculations to variations of the 
parameter space.}
\label{fig13}
\end{figure}
 
\end{document}